\documentclass[a4paper,11pt]{article}

\usepackage{amsfonts,amssymb}
\usepackage{theorem}
\usepackage{epsf}

\def\G{\Gamma}

\begin{document}

\rightline{IFUM-840-FT}

\vskip 0.8 truecm
\Large
\bf
\centerline{Weak Power-Counting Theorem for the Renormalization}
\centerline{of the Nonlinear Sigma Model in Four Dimensions}
\normalsize \rm

\normalsize
\rm
\vskip 0.5 truecm
\large
\centerline{Ruggero Ferrari 
\footnote{E-mail address: {\tt ruggero.ferrari@mi.infn.it}} 
and Andrea Quadri \footnote{E-mail address: 
{\tt andrea.quadri@mi.infn.it}}}

\vskip 0.3 truecm
\normalsize
\centerline{Phys. Dept. University of Milan, 
via Celoria 16, 20133 Milan, Italy } 
\centerline{and I.N.F.N., sezione di Milano} 

\vskip 0.7  truecm
\normalsize
\bf
\centerline{Abstract}
\rm
\begin{quotation}
\noindent
The formulation of the non linear $\sigma$-model 
in terms of flat connection
allows the construction of a perturbative solution of a local
functional equation by means of cohomological techniques which
are implemented in gauge theories.
In this paper we discuss some properties of the solution
at the one-loop level 
in $D=4$.
We prove the validity of a weak power-counting
theorem in the following form: 
although the number of divergent
amplitudes is infinite only a finite number of counterterms parameters
have to be introduced in the effective action 
in order to make the theory finite at one loop,
while respecting the functional equation (fully symmetric subtraction
in the cohomological sense). The proof uses the linearized functional
equation of which we provide the general solution in terms of
local functionals. The counterterms are expressed in terms of linear
combinations of these invariants and the coefficients are
fixed by a finite number of divergent amplitudes. 
These latter amplitudes contain only insertions of the composite operators
$\phi_0$ (the constraint of the non linear $\sigma$-model) 
and $F_\mu$ (the flat connection). 
The structure of the functional equation suggests a hierarchy 
of the Green functions. In particular once the amplitudes for the composite
operators $\phi_0$ and $F_\mu$ are given all the others can be 
derived by functional derivatives. In this paper we show that
at one loop the renormalization of the theory is achieved 
by the subtraction of divergences of the amplitudes at the top
of the hierarchy.
As an example we derive the counterterms for the four-point
amplitudes.
\end{quotation}

\begin{flushleft}
\end{flushleft}
\newpage

\section{Introduction}
Since a long time people realized that the nonlinear
$\sigma$-model cannot be renormalized in a symmetric way
by imposing global chiral symmetry
already at one loop \cite{Ecker:1972bm,Appelquist:1980ae,Tataru:1975ys}.
Some of the unwanted (chiral breaking) terms can be disposed of
by redefinition of the field (quartic divergences)
\cite{Tataru:1975ys,Gerstein:1971fm,Charap:1970xj,Honerkamp:1996va}.
However some divergent terms of the one-loop off-shell pion-pion
scattering amplitude still violate chiral symmetry and can be
reabsorbed by redefinition of the field only if 
derivatives are allowed \cite{Appelquist:1980ae}.
This strategy of removing the divergences never turned to a consistent
program both for technical difficulty and for the impossibility of
fixing the necessary finite subtractions.
>From these previous experiences it is clear that the renormalization
of the nonlinear $\sigma$-model cannot be achieved by using
chiral-invariant counterterms only.
In particular one has to find a technique to implement
the idea of field redefinition.
This problem turns out to be closely related to the issue
of identifying the good symmetry of the theory, i.e. the one that 
survives quantization.

We discuss here a unified  solution \cite{Ferrari:2005ii} 
to both problems 
which makes use of a single scalar external source
coupled to the constrained $\phi_0$ field.
The introduction of the composite operator $\phi_0$ turns out
to be unavoidable in order to discuss the implementation of
chiral symmetry at the quantum level by means
of the Ward-Takahashi identities.

Let us briefly outline the formalism by which we renormalize 
at one loop
the nonlinear $\sigma$ model.
We consider 
\cite{Ferrari:2005ii} the scalar fields ($\phi_a$) as parameters
of a flat connection (gauge field with zero field strength).
 A local functional equation encoding the underlying local invariance
property of the Haar measure in the path-integral
\begin{eqnarray}
\delta \phi_a(x) = \frac{1}{2} \alpha_a(x) \phi_0(x) + \frac{g}{2} \epsilon_{abc} \phi_b(x) \alpha_c(x) \, , ~~~~ \delta \phi_0(x) = - \frac{g^2}{2} \alpha_a(x) \phi_a(x) 
\label{intro.1}
\end{eqnarray}
is then derived.  Quantization is performed
by imposing the functional equation on the 1-PI vertex
functional in $D$-dimensions.
The functional equation embodies the relevant symmetry of the 
full quantum theory .

The projection on the physical value $D \rightarrow 4$
requires a recursive subtraction procedure of the poles.
The subtraction is implemented by a set of counterterms in the
Feynman rules in such a way to respect the functional equation.

The counterterms 
are determined by exploiting a hierarchy inherent to the 
solutions of the functional equation. Since
the counterterms have to be local functionals, 
the analysis of the functional equation can be limited
to local solutions.

In this paper we provide a general classification of the
local  functionals which are solutions of the linearized
equation.
This is the relevant equation for the counterterms 
at one loop level. Moreover it is the equation
which controls all possible finite subtractions in the dimensional
renormalization scheme. 

The invariants are integrated formal power series of local monomials
in the pion fields $\phi_a$, the external source $J_{a\mu}$
of the flat connection $F^\mu_a$ and the external source $K_0$ of the composite operator
$\phi_0$ (the constraint in the nonlinear $\sigma$-model).
They can be classified by cohomological methods implemented in gauge theories.
This provides a useful insight into the underlying geometry of the quantum
nonlinear $\sigma$ model in $D=4$.

The solution is governed by a weak power-counting theorem:
although an infinite number of divergent amplitudes
exists at one loop-level, only a finite number of them
has to be evaluated in order to make the theory finite 
at one loop level while respecting the functional equation
(fully symmetric subtraction in the cohomological sense). 
They correspond to amplitudes involving only the insertions
of the composite operators $\phi_0$ and $F^\mu_a$, i.e.
the amplitudes obtained by functional differentiation
of the 1-PI vertex functional w.r.t. $K_0$'s and $J^\mu$'s.
These amplitudes are at the top of the hierarchy implied
by the functional equation (ancestor amplitudes). 
They allow to fix uniquely
the coefficients of the invariants entering in the solution 
which parameterizes
the counterterms.

This is an 
extremely 
powerful tool for dealing with the
intricacies of divergences of the nonlinear $\sigma$-model
in $D=4$,
since all the other counterterms
(i.e. those involving at least one $\phi$ field)
can be derived
from this solution by projection on the relevant monomials.
We stress that when expanded on the basis of monomials in $\phi$'s and the
external sources the solution contains an infinite number of terms,
associated with the divergences of amplitudes with an arbitrarily
high number of pion legs. All of them are needed in order to perform 
the one-loop renormalization of the model. It is a remarkable fact
that they can be rewritten in terms of a finite number of invariants
controlled by a finite number of independent coefficients.

As an example 
we obtain the counterterms for the
set of four-point amplitudes.
Moreover we apply the method to prove a simple criterion
establishing the convergence
of amplitudes which are divergent by naive power-counting
but whose convergence is implied by the local functional equation.

This work is part of a program aiming to provide finite
Feynman amplitudes at every order in the loop expansion
of the nonlinear $\sigma$ model in $D=4$ in a symmetric scheme.
The phenomenological implications of this subtraction strategy
remain an open problem since at every order in $\hbar$ 
there is a new finite set of independent parameters
associated to in principle admissible local
counterterms.
This aspect is shared by other approaches typically focused
on the problem of giving a meaning to the loop corrections
in chiral Lagrangian models \cite{Gasser:1984gg}-\cite{Bijnens:1999hw}.

\medskip
The paper is organized as follows.
In Sect.~\ref{sec1} we describe the subtraction procedure
and the inherent weak power-counting theorem.
In Sect.~\ref{sec3} we set up the cohomological framework
needed to classify the local solutions of the linearized
functional equation. The most general local solution is
 characterized in Sect.~\ref{sec4}.
Sect.~\ref{sec5} is devoted to the parameterization
of the one-loop divergences 
in $D=4$
in terms of local invariant solutions.
As an application the counterterms of four-point amplitudes
are derived in Sect.~\ref{sec6}.
In Sect.~\ref{sec.new} we provide a comparison with similar results
obtained in chiral lagrangian theories.
Conclusions are given in Sect.~\ref{sec7}. Appendix~\ref{appA}
finally contains  a derivation of the weak-power-counting formula.

\section{Subtraction procedure}\label{sec1}

In this section we deal with the nonlinear $\sigma$-model
in the formulation given by the functional equation 
\cite{Ferrari:2005ii} which one derives from the local gauge transformations
on the associated flat connection 
\begin{eqnarray}
&& F_\mu = \frac{i}{g} \Omega \partial_\mu \Omega^\dagger = 
\frac{1}{2} F_{a\mu} \tau^a \, , \nonumber \\
&& \Omega = \frac{1}{m_D} (\phi_0 + i g \tau^a \phi_a) \, , ~~~~
\Omega^\dagger \Omega=1\, , ~~~~ {\rm det} ~~ \Omega = 1 \, , ~~~~ 
\phi_0^2 + g^2 \phi_a^2 = m_D^2 \, . \nonumber \\
\label{sec1:1}
\end{eqnarray}
$\tau^a$ are the Pauli matrices and $m_D = m^{D/2-1}$. $m$ is the mass
scale of the theory.

The local transformations are
\begin{eqnarray}
&& \Omega ' = U \Omega \, , \nonumber \\
&& F'_\mu = U F_\mu U^\dagger + \frac{i}{g} U \partial_\mu U^\dagger \, .
\label{sec1:2}
\end{eqnarray}
The local functional equation for the 1-PI generating functional 
follows from the standard path-integral formulation by using the classical
action in $D$ dimensions
\begin{eqnarray}
\G^{(0)} = \int d^Dx \, 
\Big ( \frac{1}{2} \partial_\mu \phi_a \partial^\mu \phi_a 
      + \frac{1}{2} g^2 \frac{\phi_a \partial_\mu \phi_a \phi_b \partial^\mu \phi_b}{\phi_0^2} + K_0 \phi_0 + J_{a\mu} F^\mu_a \Big ) \, .
\label{sec1:3}
\end{eqnarray}
By exploiting the invariance of the Haar measure in the path-integral
under the local gauge transformations one obtains
\begin{eqnarray}
\Big ( \frac{m_D^2}{2} \partial^\mu \frac{\delta \G}{\delta J^\mu_a}
+ g^2 K_0 \phi_a + 
\frac{\delta \G}{\delta K_0} \frac{\delta \G}{\delta \phi_a}
+ g \epsilon_{abc} \frac{\delta \G}{\delta \phi_b} \phi_c 
+ 2 D [ \frac{\delta \G}{\delta J} ]_{ab}^\mu J_{b\mu} \Big ) (x) = 0
\label{sec1:4}
\end{eqnarray}
with
\begin{eqnarray}
D[X]_{ab}^\mu = \partial^\mu \delta_{ab} - g \epsilon_{abc} X_c^\mu \, .
\label{sec1:5}
\end{eqnarray}

In order to construct the perturbative series we notice that $\G^{(0)}$
in eq.(\ref{sec1:3}) is a solution to eq.(\ref{sec1:4}) and therefore
we can read immediately from eq.(\ref{sec1:3}) the Feynman rules.

The 1-PI generating functional obtained from these rules
is a solution to eq.(\ref{sec1:4}) in $D$ dimensions.
The projection of the $D$-dimensional solution  
on the physical value $D \rightarrow 4$ requires a recursive 
subtraction procedure.

The subtraction procedure follows the hierarchy implied by eq.(\ref{sec1:4}).
This means that we fix at first the counterterms for the amplitudes
involving only the composite operators $F^\mu_a$ and $\phi_0$
(derivatives of $\G$ only w.r.t. $J^{\mu_1}_{a_1}, \dots,  J^{\mu_n}_{a_n},$ 
$\dots, K_0,\dots$).
A simple dimensional analysis indicates that the removal of the poles
in $D=4$ has to be done on the Laurent expansion of the normalized
amplitude
\begin{eqnarray}
\Big ( \frac{m_D}{m} \Big )^{2(n-1)} \G_{J^{\mu_1}_{a_1} \dots
  J^{\mu_n}_{a_n}} \, .
\label{sec1:6}
\end{eqnarray}

Eq.(\ref{sec1:4}) then constrains the correct factor for the amplitudes
involving the fields $\phi_a$ and the composite operator $\phi_0$.

We denote by $\G_{\rm pol}^{(n)}$ the corresponding pole part
of the Laurent expansion of the $n$-th order
vertex functional $\G^{(n)}$.

Our conjecture is that order by order we can modify the Feynman rules
by adding the counterterms required by dimensional subtraction,
in such a way that eq.(\ref{sec1:4}) is satisfied (symmetric subtraction).
At one loop level the removal of the pole part of the divergent
amplitudes is by means of a solution of the
linearized equation
\begin{eqnarray}
\!\!\!\!\!\!
 {\cal S}_a (\G_{\rm pol}^{(1)}) & = & 
\Big ( \frac{m_D^2}{2} \partial^\mu \frac{\delta \G_{\rm pol}^{(1)}}{\delta J^{\mu}_a}
- 2 g \epsilon_{abc} \frac{\delta \G_{\rm pol}^{(1)}}{\delta J_{c\mu}} J^\mu_b
+ \frac{\delta \G^{(0)}}{\delta K_0} \frac{\delta \G_{\rm pol}^{(1)}}{\delta \phi_a}
+ \frac{\delta \G_{\rm pol}^{(1)}}{\delta K_0} \frac{\delta \G^{(0)}}{\delta \phi_a}
\nonumber \\
& & 
+ g \epsilon_{abc} \frac{\delta \G_{\rm pol}^{(1)}}{\delta \phi_b} \phi_c 
\Big ) (x) = 0 \, 
\label{sec1:7}
\end{eqnarray}
since at this order eq.(\ref{sec1:4}) coincides with the linearized equation (\ref{sec1:7}).

The study of the solutions of eq.(\ref{sec1:7}) in terms of
local functionals provides a necessary tool in order to make consistent
the subtraction procedure outlined above.
Their coefficients have to be chosen in such a way to remove the pole parts 
of the $D$-dimensional amplitudes.

As will be shown, these coefficients are uniquely fixed
by the pole part of the divergent amplitudes which only involve the composite
operators $F_{a\mu}$ and $\phi_0$ (i.e. 1-PI Green functions
obtained by differentiating $\G$ w.r.t. the sources $J_{a\mu}$ and $K_0$).

At each order $n$ in the loop expansion only a finite number of them
exists. There is indeed a weak power-counting for the external sources
$J_{a\mu}$ and $K_0$. 
A $n$-loop graph with $N_J$ insertions of the composite operator $F_{a\mu}$,
$N_{K_0}$ insertions of the composite operator $\phi_0$ 
and no $\phi$ external legs is superficially
convergent provided that
\begin{eqnarray}
N_J + 2 N_{K_0} > (D-2)n+2 \, .
\label{sec1:8}
\end{eqnarray}
The derivation of the above formula is given in Appendix ~\ref{appA}.
Eq.(\ref{sec1:8}) fixes the upper bound on the number of independent
ancestor amplitudes.

The solutions of eq.(\ref{sec1:7}) will be given in terms of linear
combinations of invariant local functionals. The coefficients
of these invariants are in principle free parameters
and they are constrained by the functional equation (\ref{sec1:4}).
The hierarchical structure of this equation might reduce drastically
the number of independent divergent amplitudes to be evaluated.
The simplest example of this is provided by the one-loop corrections
where only the monomials in $J$ and $K_0$ and their derivatives
(present in the invariant solution)
need to be computed in terms of the pole part of the amplitudes.

\section{Background formalism}\label{sec3}

In order to classify the solutions to eq.(\ref{sec1:7}) it is
convenient to introduce a set of local parameters
$\omega_a(x)$ and rewrite eq.(\ref{sec1:7}) 
in the following equivalent form
\begin{eqnarray}
\delta \G_{\rm pol}^{(n)} &\equiv & \int d^4x \, 
\Big ( -\frac{m_D^2}{4} \partial^\mu \omega_a 
\frac{\delta \G_{\rm pol}^{(n)}}{\delta J^\mu_a}
- g \epsilon_{abc} \omega_a J^\mu_b \frac{\delta \G_{\rm pol}^{(n)}}{\delta J_{c\mu}}
\nonumber \\
& & +
\Big ( \frac{\omega_a}{2} \frac{\delta \G^{(0)}}{\delta K_0} 
+ \frac{g}{2} \epsilon_{abc} \phi_b \omega_c \Big )
\frac{\delta \G_{\rm pol}^{(n)}}{\delta \phi_a}
+ \frac{\omega_a}{2} \frac{\delta \G^{(0)}}{\delta \phi_a}
  \frac{\delta \G_{\rm pol}^{(n)}}{\delta K_0} 
\Big ) = 0
\label{bkg.1}
\end{eqnarray}

The geometrical meaning of the above equation becomes clear
after the rescaling 
\begin{eqnarray}
\tilde J_{a\mu} = -\frac{4}{m_D^2}J_{a\mu} \, .
\label{bkg.2}
\end{eqnarray}
$\tilde J_{a\mu}$ transforms as a (background) gauge connection
under the action of $\delta$
while $\Omega = \frac{1}{m_D} (\phi_0 + i g \tau^a \phi_a )$ transforms
in the fundamental representation.
For later use we notice that the transformation of $K_0$ is 
proportional to the classical equation of motion for $\phi_a$.

\medskip
There is a BRST differential $s$ \cite{Becchi:1974md, Becchi:1975nq,Piguet:1995er}
associated with the transformation in eq.(\ref{bkg.1}).
It is obtained by promoting the parameters $\omega_a$ to
classical local anticommuting parameters.
Global chiral symmetry has been discussed in a similar fashion
with the use of constant ghosts in \cite{Blasi:1986sm}.
The action of $s$ on $\tilde J_{a\mu}, \phi_a$ and $K_0$
is induced by the action of $\delta$, i.e.
\begin{eqnarray}
&& 
\!\!\!\!\!\!\!\!\!\!\!\!\!\!
s \tilde J_{a\mu} = \partial_\mu \omega_a + g \epsilon_{abc} \tilde J_{b\mu} \omega_c
\, , ~~~
   s \phi_a = \frac{1}{2} \omega_a \phi_0 + \frac{1}{2} g
   \epsilon_{abc}\phi_b\omega_c \, , ~~~ s K_0 = \frac{1}{2} \omega_a
   \frac{\delta \G^{(0)}}{\delta \phi_a} \, . \nonumber \\
\label{bkg.2bis}
\end{eqnarray}
The operator $s$
becomes nilpotent provided that we extend its action to $\omega_a$ by setting
\begin{eqnarray}
s \omega_a = -\frac{1}{2} g \epsilon_{abc} \omega_b \omega_c \, .
\label{bkg.3}
\end{eqnarray}
A conserved Faddeev-Popov ($\Phi\Pi$) charge can be introduced by 
requiring that all variables with the exception of $\omega_a$
are $\Phi\Pi$-neutral and $\Phi\Pi(\omega_a)=1$.

Eq.(\ref{bkg.1}) is equivalent to
\begin{eqnarray}
s \G_{\rm pol}^{(n)} = 0 
\label{bkg.4}
\end{eqnarray}
since there are no variables with negative $\Phi\Pi$-charge
(thus forbidding $s$-exact solutions $Y^{(n)} = s X^{(n)}$, where
$X^{(n)}$ has $\Phi\Pi$-charge $-1$, which automatically fulfill
$s Y^{(n)} = 0$ by the nilpotency of $s$).

The advantage of the BRST formulation of the local functional
equation provided by eq.(\ref{bkg.4}) is that it allows
to make use of the cohomological techniques implemented in gauge
theories \cite{Piguet:1995er},
\cite{Barnich:2000zw}-\cite{Gomis:1994he}
in order to derive an exhaustive classification of the solutions.
%

\section{Solutions of the linearized functional equation}\label{sec4}

We now move to the study of eq.(\ref{bkg.4}).
The recursive subtraction of the poles is implemented by a set of
counterterms in the Feynman rules. It is required that they are local
functionals solution of eq.(\ref{bkg.4}).

For renormalizable theories the power-counting theorem puts
dimensionality bounds on them and so this limits the number of independent
monomials.
For non-renormalizable theories as the one we are dealing with
this constraint on the number is no more present. 

On general grounds the required counterterms might in some cases reduce
to a polynomial if the perturbative expansion is cut to a finite loop order.
We will show that this is not the case 
for the nonlinear $\sigma$-model
even at one loop level: there exist divergent amplitudes involving
any number of $\phi$'s.

This apparently wild behavior is tamed by an extremely powerful hierarchy
when eq.(\ref{bkg.4}) is used in order to parameterize the
one-loop divergences.
Indeed it turns out that the counterterms are controlled 
by a linear combination of a finite
number of invariants which are solutions to eq.(\ref{bkg.4}),
as a consequence of the weak power-counting on $K_0$ and $J_{a\mu}$.
Once the relevant linear combination is known, 
all the divergences for amplitudes
involving any number of $\phi$'s and external sources 
are obtained by projection on the relevant monomial in
$\phi$'s, $K_0$ and $J_{a\mu}$.
Eq.(\ref{bkg.4}) thus provides an extremely powerful and efficient 
tool for the classification of the UV divergences in the model at hand.

In order to exploit eq.(\ref{bkg.4})
 we first need to find the most general solution to eq.(\ref{bkg.4})
in the space of integrated local functionals 
(in the sense of local formal power series)
spanned by $\phi_a$, $K_0,J_{a\mu}$ and their derivatives.
This amounts to characterize the cohomology of the nilpotent differential $s$
in eq.(\ref{bkg.2bis}) in the sector of $\Phi\Pi$-neutral local
functionals.

The required solution can be found rather easily by noticing that
the following combination
\begin{eqnarray}
{\overline{K}}_0 = \frac{m_D^2 K_0}{\phi_0} - \phi_a \frac{\delta S_0}{\delta \phi_a}
\label{sol.1}
\end{eqnarray}
is $s$-invariant. In the above equation we have set
\begin{eqnarray}
S_0 = \frac{m_D^2}{8} \int d^Dx \, \Big ( F_{a\mu} + \frac{4}{m_D^2} J_{a\mu} \Big )^2 \, .
\label{sol.2}
\end{eqnarray}

By exploiting the invariance of $S_0$ under $s$ we obtain
\begin{eqnarray}
s \overline{K}_0 & = & \frac{m_D^2}{2 \phi_0}  \omega_a 
\frac{\delta \G^{(0)}}{\delta \phi_a} + \frac{g^2 m_D^2}{2 \phi_0^2} K_0 ~ \omega_a \phi_a
+ \Big [ s, - \phi_a \frac{\delta}{\delta \phi_a} \Big ] S_0 \nonumber \\
& = & \frac{m_D^2}{2 \phi_0}  \omega_a 
\frac{\delta S_0}{\delta \phi_a} 
- \frac{g^2 m_D^2}{2 \phi_0^2} K_0 ~ \omega_a \phi_a
+ \frac{g^2 m_D^2}{2 \phi_0^2} K_0 ~ \omega_a \phi_a \nonumber \\
& & + \Big [ s, - \phi_a \frac{\delta}{\delta \phi_a} \Big ] S_0 \nonumber \\
& = & \frac{m_D^2}{2 \phi_0}  \omega_a 
\frac{\delta S_0}{\delta \phi_a} 
+ \Big [ s, - \phi_a \frac{\delta}{\delta \phi_a} \Big ] S_0 \, .
\label{calc.1}
\end{eqnarray}
By taking into account that $S_0$ does not depend on $K_0$ we also get
\begin{eqnarray}
\Big [ s, - \phi_a \frac{\delta}{\delta \phi_a} \Big ] S_0 & = & 
- \frac{1}{2} \omega_a \phi_0 \frac{\delta S_0}{\delta \phi_a}
- \frac{g^2}{2 \phi_0} \omega_a \phi_b^2 \frac{\delta S_0}{\delta \phi_a} \, .
\label{calc.2}
\end{eqnarray}
Use of eq.(\ref{calc.2}) into eq.(\ref{calc.1}) yields finally
\begin{eqnarray}
s \overline{K}_0 & = &
\frac{1}{2 \phi_0}  \omega_a ( m_D^2 - g^2 \phi_b^2) 
\frac{\delta S_0}{\delta \phi_a} 
- \frac{1}{2} \omega_a \phi_0 \frac{\delta S_0}{\delta \phi_a}
\nonumber \\
& = & \frac{1}{2} \omega_a  \phi_0
\frac{\delta S_0}{\delta \phi_a} 
- \frac{1}{2} \omega_a \phi_0 \frac{\delta S_0}{\delta \phi_a} = 0
\label{calc.3}
\end{eqnarray}
where use has been made of the last of eqs.(\ref{sec1:1}).

Since the transformation in eq.(\ref{sol.1}) is invertible
we can change variables and use $\phi_a$, $J_{a\mu}$ and $\overline{K}_0$.
$\overline{K}_0$ is invariant under $s$ while the $s$-variation of
$\phi_a$ and $J_{a\mu}$ does not contain $\overline{K}_0$.

Hence 
the computation of the cohomology of $s$ 
in eq.(\ref{bkg.2bis}) in the $\Phi\Pi$-neutral sector reduces to 
that of the BRST differential for the gauge group $SU(2)$ 
(non-linearly
represented on the group element $\Omega$) in the space of local
functionals with zero $\Phi\Pi$-charge.
This is easily seen by identifying  the $SU(2)$ connection with  $\tilde J_{a\mu}$ in eq.(\ref{bkg.2}),
while $\phi_a$ are the parameters controlling the non-linear
representation of the gauge group by the matrix $\Omega$.
$\overline{K}_0$ is an additional variable which does not transform
under $s$.

The cohomology of the BRST differential for non-linear
representations of the gauge group $SU(2)$
is known in full generality
\cite{Barnich:2000zw,Henneaux:1998hq}.
This allows us to state the following

\medskip
{\bf Proposition}. The most general local solution to eq.(\ref{bkg.4}) is
an integrated BRST (eq.(\ref{bkg.2bis}))-invariant local formal power series constructed from 
the invariant combination $\overline{K}_0$ and its ordinary
derivatives, the undifferentiated group element $\Omega$ and
the combination $F^\mu_a + \frac{4}{m_D^2} J^\mu_a$ and its subsequent covariant derivatives
w.r.t. $F_\mu$.

\medskip
The proof of this result is based on cohomological techniques and
is detailed in \cite{Barnich:2000zw,Henneaux:1998hq}.
Here we only wish to make a few comments. 

The combination
\begin{eqnarray}
F_a^\mu + \frac{4}{m_D^2} J_a^\mu = F^\mu_a - \tilde J_a^\mu
\label{comm.1}
\end{eqnarray}
is the difference of two $SU(2)$ connections and thus it 
transforms in the adjoint representation of $SU(2)$:
\begin{eqnarray}
s \Big ( F^\mu_a - \tilde J_a^\mu \Big ) = g \epsilon_{abc} 
( F^\mu_b - \tilde J_b^\mu) \omega_c \, .
\label{comm.2}
\end{eqnarray}
Moreover we notice that
covariant derivatives have to be understood only w.r.t. $F_\mu$.
Covariant derivatives w.r.t. $J_\mu$ can also be used in order to construct invariants.
However these invariants are not independent, since a covariant derivative
w.r.t. $J^\mu$ can be replaced by a covariant derivative
w.r.t. $F^\mu$ plus a term containing the combination $F^\mu + \frac{4}{m_D^2} J^\mu$.

Finally in the sector with at least one derivative there is still
the freedom to perform an integration by parts in order to reduce
the number of independent invariants.   
Once this ambiguitiy is taken into account one gets the set
of independent invariants  on which to project the solutions to
eq.(\ref{bkg.4}).

The above Proposition is a very powerful result allowing for a simple constructive
characterization of the solutions to eq.(\ref{bkg.4}). In the next
section we will show how to make use of it in order 
to specify completely the whole set of one-loop counterterms.

\section{One-loop counterterms}\label{sec5}

>From the above discussion we can deal with the one-loop corrections
in $D=4$
by writing the most general local solution to eq.(\ref{bkg.4})
compatible with the weak power-counting. Since eq.(\ref{bkg.4}) is linear,
the solution is a linear combination of the following invariants
(all covariant derivatives are understood w.r.t. the flat connection
$F_\mu$):
\begin{eqnarray}
&& {\cal I}_1 = \int d^Dx \, \Big [ D_\mu ( F + \frac{4}{m_D^2} J )_\nu \Big ]_a \Big [ D^\mu ( F + \frac{4}{m_D^2} J )^\nu \Big ]_a  \, , 
\nonumber \\
&& {\cal I}_2 = \int d^Dx \, \Big [ D_\mu ( F + \frac{4}{m_D^2} J )^\mu \Big ]_a \Big [ D_\nu ( F + \frac{4}{m_D^2} J )^\nu \Big ]_a  \, , 
\nonumber \\
&& {\cal I}_3 = \int d^Dx \, \epsilon_{abc} \Big [ D_\mu ( F + \frac{4}{m_D^2} J )_\nu \Big ]_a \Big ( F + \frac{4}{m_D^2} J \Big )^\mu_b \Big ( F + \frac{4}{m_D^2} J \Big )^\nu_c \, ,  \nonumber \\
&& {\cal I}_4 = \int d^Dx \, \Big ( \frac{m_D^2 K_0}{\phi_0} - \phi_a \frac{\delta S_0}{\delta \phi_a} \Big )^2 \, , \nonumber \\
&& {\cal I}_5 = \int d^Dx \, \Big ( \frac{m_D^2 K_0}{\phi_0} - \phi_a \frac{\delta S_0}{\delta \phi_a} \Big ) \Big ( F + \frac{4}{m_D^2} J \Big )^2 \, , 
\nonumber \\
&& {\cal I}_6 = \int d^Dx \, \Big ( F + \frac{4}{m_D^2} J \Big  )^2
 \Big ( F + \frac{4}{m_D^2} J \Big )^2 \, , \nonumber \\
&& {\cal I}_7 = \int d^Dx \, \Big ( F + \frac{4}{m_D^2} J \Big  )^\mu_a
   \Big ( F + \frac{4}{m_D^2} J\Big  )^\nu_a \nonumber \\
&& ~~~~~~~~~~~~~~~~
   \Big ( F + \frac{4}{m_D^2} J \Big  )_{b\mu}
   \Big ( F + \frac{4}{m_D^2} J \Big  )_{b\nu} \, . 
\label{ex.1}
\end{eqnarray}

\medskip
A few comments on this list are in order. ${\cal I}_1$ and ${\cal I}_2$
describe the pole part of the 2-point function $\G^{(1)}_{JJ}$.
${\cal I}_3$ is the only invariant that can yield the counterterm
associated with $\G^{(1)}_{JJJ}$.
Finally ${\cal I}_6$ and ${\cal I}_7$ control the pole part of the 
4-point function
$\G^{(1)}_{JJJJ}$, while
the 2-point function $\G^{(1)}_{K_0K_0}$
and the 3-point function $\G^{(1)}_{K_0 JJ}$ are related
to ${\cal I}_4$ and ${\cal I}_5$.
We notice that the functional equation in eq.(\ref{sec1:4}) allows
to derive $\G^{(1)}_{K_0 K_0}$ and
$\G^{(1)}_{K_0 JJ}$ from $\G^{(1)}_{JJJJ}$, $\G^{(1)}_{JJJ}$ and
$\G^{(1)}_{JJ}$. Therefore only three amplitudes have to be
computed.

The correct linear combination of the invariants has to be
found by comparison with the solution of eq.(\ref{sec1:4})
which is valid in $D$-dimensions. Therefore the coefficients
must contain the correct power of $m_D$. Once
these coefficients have been established all the one-loop
divergences for amplitudes involving any number of $\phi$'s are described
by the projection of the solution on the relevant monomial.
In fact all the amplitudes involving at least one $\phi$ field
can be derived by subsequent use of the functional equation
(\ref{sec1:4}).

We denote by $\hat \G^{(1)} = - \G^{(1)}_{\rm pol}$ 
the one-loop divergent counterterms.

By direct computation one finds $\hat \G^{(1)}[JJ]$ and $\hat \G^{(1)}[JJJ]$
\cite{Ferrari:2005ii}
\begin{eqnarray}
&& \hat \G^{(1)}[JJ] = \frac{1}{D-4} \Big ( \frac{m}{m_D} \Big )^2
\frac{g^2}{12 \pi^2 m^4} \int d^Dx \, J_a^\mu (\square g_{\mu\nu} 
- \partial_\mu \partial_\nu) J^\nu_a \, , \nonumber \\
&& \hat \G^{(1)}[JJJ] = \frac{1}{D-4} 
\frac{1}{3\pi^2} \Big ( \frac{g}{m^2} \Big )^3
\Big ( \frac{m}{m_D} \Big )^4 \int d^Dx \, \epsilon_{abc}
\partial_\mu J_{a\nu} J^\mu_b J^\nu_c \, .
\label{ex.2}
\end{eqnarray}
This fixes the coefficients of ${\cal I}_1, {\cal I}_2, {\cal I}_3$
which enter into the solution in the combination
\begin{eqnarray}
- \frac{1}{D-4} \frac{1}{12} \frac{g^2}{(4\pi)^2} \frac{m_D^2}{m^2} \Big (
{\cal I}_1 - {\cal I}_2 - g {\cal I}_3 \Big ) \, .
\label{ex.3}
\end{eqnarray}
Direct computation of the pole part of $\G^{(1)}_{JJJJ}$ gives
\begin{eqnarray}
\hat \G^{(1)}[JJJJ] & = & \frac{1}{D-4} \frac{1}{3 (4\pi)^2} 
\Big ( \frac{2g}{m^2} \Big )^4
\Big ( \frac{m}{m_D} \Big )^6 \nonumber \\
&& ~~ \int d^Dx \, (J_{a\mu} J^\mu_a J_{b\nu} J^\nu_b
+ 2 J_{a\mu} J_{a\nu} J_b^\mu J^\nu_b \Big ) \, . \nonumber \\
\label{ex.4}
\end{eqnarray}
This in turn fixes the coefficients of ${\cal I}_6$ and ${\cal I}_7$
in the combination
\begin{eqnarray}
\frac{1}{D-4} \frac{1}{(4\pi)^2} \frac{g^4}{48} \frac{m_D^2}{m^2} 
\Big ( {\cal I}_6 + 2 {\cal I}_7 \Big ) \, .
\label{ex.5}
\end{eqnarray}
Finally from the counterterms
\begin{eqnarray}
\hat \G^{(1)}[K_0 K_0] = \frac{1}{D-4} \frac{3g^4}{2m^2} \frac{1}{(4\pi)^2}
\int d^Dx \, K_0^2 (x)
\label{ex.6}
\end{eqnarray}
and
\begin{eqnarray}
\hat \G^{(1)}[K_0 JJ] = 
\frac{1}{D-4}\frac{8g^4}{m^5} \frac{1}{(4\pi)^2} \Big ( \frac{m}{m_D} \Big )^3
\int d^Dx \, K_0(x) J^2(x)
\label{ex.7}
\end{eqnarray}
we get the coefficients of ${\cal I}_4$ and ${\cal I}_5$:
\begin{eqnarray}
\frac{1}{D-4} 
\frac{1}{(4\pi)^2} \frac{3}{2} \frac{g^4}{m^2 m_D^2} {\cal I}_4
+ \frac{1}{D-4} \frac{1}{(4\pi)^2} \frac{1}{2} \frac{g^4}{m^2} {\cal I}_5 \, .
\label{ex.8}
\end{eqnarray}
Therefore the full set of one-loop divergent counterterms is given by
the functional
\begin{eqnarray}
\hat \G^{(1)} & = & \frac{1}{D-4} \Big [
- \frac{1}{12} \frac{g^2}{(4\pi)^2} \frac{m_D^2}{m^2} \Big (
{\cal I}_1 - {\cal I}_2 - g {\cal I}_3 \Big ) 
+  \frac{1}{(4\pi)^2} \frac{g^4}{48} \frac{m_D^2}{m^2} 
\Big ( {\cal I}_6 + 2 {\cal I}_7 \Big ) \nonumber \\
& & + 
\frac{1}{(4\pi)^2} \frac{3}{2} \frac{g^4}{m^2 m_D^2} {\cal I}_4
+ \frac{1}{(4\pi)^2} \frac{1}{2} \frac{g^4}{m^2} {\cal I}_5 \Big ] \, .
\label{ex.9}
\end{eqnarray}
These are the counterterms to be used in $D$-dimensional perturbation
theory. This is the reason why $m_D$ is put in evidence. Moreover the presence
of $m_D$ both in the coefficients and the invariants fixes non-trivial
finite parts of the counterterms beyond the pole part in $\frac{1}{D-4}$.
These finite parts are non-trivial since they are needed to maintain
the validity of the functional equation after subtraction.

Eq.(\ref{ex.9}) is not the most general solution. One can always
add finite solutions of $sX =0$. It is a choice that we make in this
paper to perform a minimal subtraction on the basis of simplicity and
elegance.

The explicit form of the counterterms (\ref{ex.9}) allows us to comment on two
further important points.

One is the issue of chiral invariance of the counterterms at one loop.
By direct inspection one sees that, after putting $J_\mu^a = K_0 =0$,
${\cal I}_1, {\cal I}_2, {\cal I}_3, {\cal I}_6$ and ${\cal I}_7$ are
chiral invariant (global transformation) while both ${\cal I}_4$ and
${\cal I}_5$ are not chiral invariant.
Therefore the counterterms at one loop do not maintain chiral invariance
as noted in \cite{Ecker:1972bm,Appelquist:1980ae,Tataru:1975ys}.

As a last point eq.(\ref{ex.9}) accounts for the fact that
the chiral-breaking counterterms are associated to the
renormalization of the insertion of the composite operator
$\phi_0$ coupled to the source $K_0$.

\section{Examples}\label{sec6}

The use of eq.(\ref{ex.9}) is straightforward. One needs only to perform
the relevant functional derivatives of the local functional
$\hat \G^{(1)}$.

As an example we can get the counterterm for the four-point function
by projecting $\hat \G^{(1)}$ in eq.(\ref{ex.9})
on the monomials involving $\phi_a$, $K_0$ and $J_{a\mu}$.
First we consider the four-point function of the scalar fields.
By direct computation the projection of the combination
${\cal I}_1 - {\cal I}_2 - g {\cal I}_3$ on the relevant monomials
is zero, while the contribution from ${\cal I}_6 + 2 {\cal I}_7$
and ${\cal I}_4$, ${\cal I}_5$ gives rise to
\begin{eqnarray}
\!\!\!\!\!\!\!
\hat \G^{(1)} [\phi\phi\phi\phi] & = & - \frac{1}{D-4}
\frac{g^4}{m_D^2 m^2 (4 \pi)^2} \nonumber \\
&& ~~~ \int d^D x \, 
\Big [ 
- \frac{1}{3} \partial_\mu \phi_a \partial^\mu \phi_a  
\partial_\nu \phi_b \partial^\nu \phi_b 
- \frac{2}{3} \partial_\mu \phi_a \partial_\nu \phi_a 
              \partial^\mu \phi_b \partial^\nu \phi_b  
\nonumber \\
&& ~~~~
-\frac{3}{2} \phi_a \square \phi_a \phi_b \square \phi_b - 2 \phi_a \square \phi_a \partial_\mu \phi_b \partial^\mu \phi_b
\Big ] \, .
\label{ex.10}
\end{eqnarray}
The terms in the first line between square brackets 
are associated to global chiral-invariant
counterterms \cite{Appelquist:1980ae,Slavnov}. They are generated by 
the combination ${\cal I}_6 + 2 {\cal I}_7$.
These invariants are constructed from the geometrical quantities
given by the  flat connection $F_\mu$ and the background connection
$\tilde J_\mu$.
The terms in the second line are obtained from the projection
of the invariants ${\cal I}_4$ and ${\cal I}_5$, which are controlled by
$\hat \G^{(1)}_{K_0K_0}$ and $\hat \G^{(1)}_{K_0 J J}$.
The latter encode the renormalization of the external source $K_0$.
In \cite{Appelquist:1980ae,Tataru:1975ys} they were obtained by means of a (non-locally invertible)
field redefinition of $\phi_a$.

We also provide the counterterms for the remaining four-point functions.
By projection on the relevant monomials we obtain
\begin{eqnarray}
\!\!\!\!\!\!\!
\hat \G^{(1)}[JJJ\phi] & = & \frac{1}{D-4} \frac{8}{(4\pi)^2} 
\frac{g^4}{m^2} \frac{1}{m_D^5} \nonumber \\
&& \int d^Dx \, 
\phi_a \Big ( 2 \partial J^a J^2 - \frac{8}{3} \partial_\nu
(J^\nu_a J^2) 
- \frac{4}{3} \partial_\nu ( J^\nu_c J^\mu_c J_{a\mu}) 
\Big ) 
\label{ex.11.1}
\end{eqnarray}
\begin{eqnarray}
\!\!\!\!\!\!\! 
\hat \G^{(1)}[JJ\phi\phi] & = & - \frac{1}{D-4}
\frac{4}{(4\pi)^2} \frac{g^4}{m^2} 
\frac{1}{m_D^4} \int d^Dx \, \Big ( 
\frac{4}{3} \partial_\mu \phi_a \partial^\mu \phi_a J^2 + \partial_\mu \phi_a^2 \partial^\mu J^2 \nonumber \\
& & + \frac{1}{2} \phi_c \phi_b \partial J_b \partial J_c 
    - \frac{4}{3} \partial^\mu \phi_c \partial_\mu \phi_b J^\nu_c J_{b\nu}
\nonumber \\
& & + \frac{8}{3} \phi_c \partial_\mu \phi_b (J^\nu_c \partial_\nu J^\mu_b
- J^\nu_b \partial_\nu J^\mu_c) + \frac{1}{2} \phi_c \phi_b
J^\nu_c \partial_\nu \partial J_b 
\nonumber \\
& & -\frac{2}{3} \partial^\mu \phi_c \phi_b ( \partial_\mu J^\nu_c J_{b\nu} 
- J^\nu_c \partial_\mu J_{b\nu}) \nonumber \\
& & - \frac{8}{3} \partial_\mu \phi_c \partial_\nu \phi_b J^\nu_c J^\mu_b
- \frac{4}{3} \partial_\mu \phi_c \partial_\nu \phi_c J^\nu_b J^\mu_b
\Big )
\label{ex.11.2}
\end{eqnarray}
\begin{eqnarray}
\hat \G^{(1)}[J\phi\phi\phi] & = & \frac{1}{D-4}
\frac{2g^4}{m^2 (4\pi)^2} \frac{1}{m_D^3} \int d^Dx \,  \nonumber \\
& & ~ \Big ( \frac{1}{2} J^\mu_a \partial_\mu \phi_a \square \phi^2
- J_a^\mu \partial_\mu \phi_a \partial_\nu \phi_d \partial^\nu \phi_d
\nonumber \\
& & ~~ + J^\mu_a \phi_a \partial_\mu ( \partial_\nu \phi_d \partial^\nu \phi_d)
- \frac{3}{2} J^\mu_a \phi_a \partial_\mu \square \phi^2 \nonumber \\
& & ~~ - \frac{2}{3} J^\mu_a \phi_c (\square g_{\mu\nu} - \partial_\mu \partial_\nu) (\partial^\nu \phi_c \phi_a) + 2 J^\mu_a \partial_\mu \phi_c \partial^\nu \phi_a \partial_\nu \phi_c \Big ) \nonumber \\
\label{ex.11.3}
\end{eqnarray}
\begin{eqnarray}
&& 
\!\!\!\!\!\!\!\!\!\!\!\!\!\!\!\!\!\!\!\!\!\!\!\!\!\!\!\!\!\!\!\!\!\!\!\!\!\!\!\!\!\!\!\!\!
\hat \G^{(1)}[K_0 K_0 \phi\phi] =  \frac{1}{D-4} \frac{1}{(4\pi)^2}
\frac{3}{2} \frac{g^6}{m^2 m_D^2} \int d^Dx \, K_0^2 \phi_a^2 
\\
&& 
\!\!\!\!\!\!\!\!\!\!\!\!\!\!\!\!\!\!\!\!\!\!\!\!\!\!\!\!\!\!\!\!\!\!\!\!\!\!\!\!\!\!\!\!\!
\hat \G^{(1)}[K_0 J \phi\phi] = - \frac{1}{D-4} \frac{1}{(4\pi)^2}
\frac{4 g^5}{m^2 m_D^3} \int d^Dx \, K_0 \epsilon_{abc} \partial_\mu \phi_b \phi_c J^\mu_a \, .
\label{ex.11.4}
\end{eqnarray}
We would like to make some additional comments on eq.(\ref{ex.9}).
First we notice that the expansion of $\hat \G^{(1)}$
on a basis of monomials in $\phi$, $K_0$, $J_\mu$ and their
derivatives contains terms of arbitrarily high order in the number
of $\phi$'s. Therefore there is an infinite set of 
divergent amplitudes involving the fields $\phi$. Nevertheless
they are all controlled by eq.(\ref{ex.9}), which contains
only a finite number of invariants.

Eq.(\ref{ex.9}) 
provides a full control on
the divergences of the theory. For instance the amplitude $\G^{(1)}_{JJJJ\phi}$ is
divergent by simple power-counting. It is convergent due
to the cancellations implied by the functional equation in eq.(\ref{sec1:4}),
as it can be explicitly checked.
This can be seen in an easier way from eq.(\ref{ex.9}) by noticing that
  the projection of $\hat \G^{(1)}$ on
$JJJJ\phi$ is zero. 

More generally 
the following simple criterion holds true: whenever the projection of
$\hat \G^{(1)}$ on some monomial is zero, the corresponding
amplitude is finite. 

\section{Comparison with chiral lagrangian theories} \label{sec.new}

In the present work we focused on the symmetric subtraction
of the divergences in the nonlinear sigma model and therefore
particular care has been put to write the most general
counterterms in $D$-dimensions 
(addressing in particular their dependence on $m_D$).
Moreover the powerful
strategy, based on the hierarchy of the functional equation, plays
a crucial role for the validity of the weak power-counting.

The counterterms obtained in eq.(\ref{ex.9}) can be compared with
a similar result in chiral lagrangian models.  In order to make
the comparison an easy task we use in this Section a set of notations
very close to the 
ones
adopted in the specialized literature on chiral
perturbation theory.

The counterterms of the chiral lagrangian will be written in terms of the
invariants ${\cal I}_1-{\cal I}_7$ by means of two quantities that
are essential in our approach: the external currents $\xi^i$ coupled
to the fields $U^i$ are introduced as a Legendre conjugate
\begin{eqnarray}
\xi^i = - \frac{\delta \G^{(0)}}{\delta U^i} \, ,
\label{comp.1}
\end{eqnarray}
and moreover the flat connection is introduced by 
\begin{eqnarray}
F_\mu = i U \partial_\mu U^\dagger = F_\mu^i \frac{\tau^i}{2}
\label{comp.2}
\end{eqnarray}
where 
\begin{eqnarray}
U = U_0 + i U^i \tau^i \, .
\label{comp.3}
\end{eqnarray}
The tree-level effective action is
\begin{eqnarray}
\G^{(0)} =\int d^4x \, \Big (  \frac{f^2}{4} {\rm Tr} (F_\mu - L_\mu)^2 + \xi^0 U^0 \Big ) \, .
\label{comp.4}
\end{eqnarray}
In this notations the $s$ operator becomes (in the zero ghost number sector)
\begin{eqnarray}
&& s = \int d^4x \, \frac{\omega^a}{2} \Big ( ( \delta^{ab} U^0 
+ \epsilon^{abc} U^c) \frac{\delta}{\delta U^b}
+ \frac{\delta \G^{(0)}}{\delta U^a} \frac{\delta}{\delta \xi^0} \nonumber \\
&& ~~~~~~~~~~~~~~~~~~
+ ( - 2 \partial^\mu \delta^{ab} + 2 \epsilon^{abc} L_\mu^c )  \frac{\delta}{\delta L_\mu^b} \Big ) \, .
\label{comp.5}
\end{eqnarray}
One gets
\begin{eqnarray}
&& s F_\mu^i = \partial_\mu \omega^i + \epsilon^{ijk} F_\mu^j \omega^k \,  ,
\nonumber \\
&& s L_\mu^i = \partial_\mu \omega^i + \epsilon^{ijk} L_\mu^j \omega^k \, .
\label{comp.6}
\end{eqnarray}
It is straightforward to find the transformation properties of $\xi^i$:
\begin{eqnarray}&&
s \xi^i  =  s \Big ( - \frac{\delta \G^{(0)} }{\delta U^i} \Big ) 
         =  - \Big [s, \frac{\delta}{\delta U^i} \Big ] \G^{(0)} - 
                \frac{\delta}{\delta U^i} (s \G^{(0)}) 
\nonumber \\&&
         =  + \frac{\omega_i}{2} \xi^0 \ - \epsilon^{iab} \omega^a \xi^b \, .
\label{comp.7}
\end{eqnarray}
Moreover
\begin{eqnarray}
s \xi^0 = \frac{\omega^a}{2} \frac{\delta \G^{(0)}}{\delta U^a} 
= - \frac{1}{2} \omega^a \xi^a \, .
\label{comp.8}
\end{eqnarray}
Therefore $(\xi^0, \xi^i)$ transform like $(U^0, U^i)$.
The transformation properties in eqs. (\ref{comp.6}), (\ref{comp.7})
and (\ref{comp.8}) allow the conctruction of invariant local
counterterms by using the covariant derivatives
\begin{eqnarray}
\nabla_\mu U \equiv (\partial_\mu -i L_\mu)U
= i (F - L)_\mu U.
\label{comp.8.1}
\end{eqnarray}
An useful relation can be obtained from the identity
\begin{eqnarray}
s~\int d^4 x{\rm Tr}\left(F - L\right)^2=0
\label{comp.8.2}
\end{eqnarray}
i.e. 
\begin{eqnarray}&&
 \frac{1}{2}( \delta^{ab} U^0 
+ \epsilon^{abc} U^c) \frac{\delta}{\delta U^b}
\int d^4 x{\rm Tr}\left(F - L\right)^2
\nonumber\\&&
= -2 (\partial^\mu \delta^{ab} -  \epsilon^{abc} L_\mu^c ) 
(F - L)^\mu_b=-2 D[L]_{ab\mu}(F - L)^\mu_b.
\label{comp.8.3}
\end{eqnarray}
The square is 
\begin{eqnarray}&&
( \delta^{bb'}-U^b U^{b'}
) \frac{\delta}{\delta U^b}
\int d^4 x{\rm Tr}\left(F - L\right)^2
\frac{\delta}{\delta U^{b'}}\int d^4 y{\rm Tr}\left(F - L\right)^2
\nonumber\\&&
=16 \left(D[L]_{ab\mu}(F - L)^\mu_b\right)^2.
\label{comp.8.4}
\end{eqnarray}
By using eq. (\ref{comp.8.4}) one gets
\begin{eqnarray}&&
\int d^4 x(\xi^0\xi^0+\vec\xi^2)= 
\int d^4 x \left (\xi^0\xi^0 + \left [- \frac{\delta}{\delta U^{b}}
\int d^4 y{\frac{f^2}{4}\rm Tr}\left(F - L\right)^2 +\xi^0 \frac{U^b}{U^0}
\right ]^2\right)
\nonumber\\&&
=
\int d^4 x \Big(\left(\frac{\xi^0}{U^0}
-U^{b}\frac{\delta}{\delta U^{b}}
\int d^4 y{\frac{f^2}{4}\rm Tr}\left(F - L\right)^2 
\right)^2 
+\frac{1}{4}\left(D[L]_{ab\mu}(F - L)^\mu_b\right)^2
\Big),
\nonumber\\
\label{comp.8.5}
\end{eqnarray}
where the last two terms can be identified as ${\cal I}_4$ and ${\cal
  I}_2$
in eq. (\ref{ex.1}).
\par
The correspondence with our conventions is obtained by setting
\begin{eqnarray}
& f = m_D \, , ~~~ g =1 \, , ~~~ U^0 = \frac{1}{m_D} \phi_0 \, , ~~~ U^i = \frac{1}{m_D} \phi_i \, , & \nonumber \\ 
& 
\xi^0 = m_D K_0 \, , ~~~ \tilde J^\mu_i = L^{\mu i} \, . &
\label{comp.9}
\end{eqnarray}
The correspondence with the notations used in \cite{Gasser:1983yg} is obtained by the following prescription
\begin{eqnarray}
& \!\!\!\!  f = F \, , ~~~
\xi^0 = F^2  \chi^0 \, , ~~~ \tilde \chi = 0 \, ,~~~  L^i_\mu  = (a_\mu^i + v_\mu^i) \, , ~~~~
a_\mu^i = v_\mu^i \, . &
\label{comp.10}
\end{eqnarray}
By using eqs.(\ref{comp.9}) and (\ref{comp.10}) we are in a position to
express the chiral invariants of \cite{Gasser:1983yg}
on the basis
given by the invariants ${\cal I}_1 - {\cal I}_7$:
\begin{eqnarray}
&& \int d^4x \,  (\nabla^\mu U^T \nabla_\mu U)^2  =   
\frac{1}{16} {\cal I}_6 \, , \nonumber \\
&& \int d^4x \,  
(\nabla^\mu U^T \nabla^\nu U) (\nabla_\mu U^T \nabla_\nu U)
 =  \frac{1}{16} {\cal I}_7  \, , 
\nonumber \\
&& \int d^4x \,
(\chi^T U)^2  =  \frac{1}{m^4}{\cal I}_4 \, , \nonumber \\
&& \int d^4x \,
(\nabla^\mu \chi^T \nabla_\mu U)  = 
\frac{1}{4 m^2} {\cal I}_5 - \frac{1}{4} {\cal I}_2 
\, , \nonumber \\
&& \int d^4x \, 
(U^T F^{\mu\nu} F_{\mu\nu} U) =   -\frac{1}{2} {\cal I}_1 + \frac{1}{2} {\cal I}_2 +  {\cal I}_3 -\frac{1}{4} {\cal I}_6 + \frac{1}{4} {\cal I}_7 \, , \nonumber \\
&& \int d^4x \, 
(\nabla^\mu U)^T F_{\mu\nu} (\nabla^\nu U)  = \frac{1}{4} {\cal I}_3+
         \frac{1}{8}  ({\cal I}_7 - {\cal I}_6) \, , \nonumber \\
&& \int d^4x \, 
 (\chi^T \chi)  =   \frac{1}{m^4} {\cal I}_4  + \frac{1}{4} {\cal I}_2\,  , \nonumber \\
&& \int d^4x \, {\rm Tr} F_{\mu\nu} F^{\mu\nu}  =  -2 {\cal I}_1 + 2 {\cal I}_2 + 4  {\cal I}_3 - {\cal I}_6 + {\cal I}_7 \, .
\label{comp.11}
\end{eqnarray}
By making use of the above correspondence table it is then easy to verify
that the divergent part of the 
counterterms obtained in \cite{Gasser:1984gg} coincide with those
given by eq.(\ref{ex.9}).

One should however realize that the $D$-dimensional counterterms 
in eq.(\ref{ex.9}) have a non-trivial dependence on $m_D$.
The latter gives rise to finite parts which are crucial
in order to maintain the validity of the functional
equation in the recursive subtraction procedure at higher orders
in the loop expansion. 
See for instance the explicit calculation at the two-loop
level in \cite{Ferrari:2005fc}.

\section{Conclusions}\label{sec7}

In this paper we have shown that at the one loop level 
the nonlinear $\sigma$-model can be renormalized by
using dimensional subtraction in such a way that the
defining functional equation is preserved.

The construction of the counterterms is based on the symmetry
property generated by a nilpotent operator $s$ which
transforms fields and external sources in a BRST fashion.
This operator is obtained as a linearized form of the functional
equation in the loop expansion.

Both the functional equation and the operator $s$ express
a hierarchy structure of the Green functions. 
The ancestors at the top are given by the Green functions involving only
the external sources of the flat connection $F_\mu$ and the constrained
field $\phi_0$.

A weak power-counting theorem then follows 
stating that, although
the number of divergent amplitudes is infinite, only a finite
number of counterterms parameters has to be introduced in the effective
action in order to make the theory finite 
at one loop level while respecting the functional equation
(fully symmetric subtraction in the cohomological sense).

The counterterms are then a linear combination of the $s$-invariants.
The weak power-counting limits the number of invariants needed for
the complete renormalization at the one-loop level. The amplitudes
involving only insertions of the composite operators $F^\mu_a$ and
$\phi_0$ uniquely fix the coefficients of the local invariants
entering in the linear combination which parameterizes 
the one-loop counterterms. 
All the remaining
 divergent amplitudes can be obtained by projection
of the linear combination on the appropriate monomials.

The structure of the counterterms reveals that both the pole parts
and the finite parts have to be properly fixed in order to
maintain the validity of the unsubtracted functional equation.
Moreover by inspection one sees that some of the counterterms
are not chiral invariant. These are associated to invariants
containing the external source of the constrained field $\phi_0$.

As an example we have derived the expressions for the counterterms
of the set of four-point functions.
Amplitudes associated with monomials which are not contained
in this linear combination are convergent (although their superficial
degree of divergence may be non-negative).

In $D=4$
 the whole structure of one-loop 
divergences of the nonlinear $\sigma$-model is determined
in terms of the finite set of invariants with given coefficients
in eq.(\ref{ex.9}).
This allows to renormalize completely the theory at one-loop order.

We emphasize that the $D$-dimensional counterterms 
in eq.(\ref{ex.9}) contain a non-trivial dependence on $m_D$.
The latter gives rise to finite parts which prove to be crucial
in order to maintain the validity of the functional
equation in the recursive subtraction procedure at higher orders
in the loop expansion.

\appendix

\medskip

\section{Weak power-counting for $J_\mu^a$ and $K_0$}\label{appA}

Let $G$ be a $n$-loop graph with $I$ internal lines and a 
certain set of vertices described by a collection of non-negative integers
\begin{eqnarray*}
&&  \{ V_J^{(2)}, V_J^{(3)}, V_J^{(5)}, \dots, V_J^{(2p+1)}, \dots, \nonumber \\
&& V_{K_0}^{(2)}, V_{K_0}^{(4)}, \dots, V_{K_0}^{(2q)}, \dots, \nonumber \\
&& V_{\phi}^{(4)},V_{\phi}^{(6)}, \dots, V_{\phi}^{(2r)}, \dots \}.
\end{eqnarray*}

$V_J^{(m)}$, $m=2$ or $m=3,5,7,
\dots$ denotes the number of vertices in $G$ with the insertion of one $J$
and $m$ $\phi$'s. $V_{K_0}^{(m)}$, $m=2,4,6,\dots$ stands for
the number of vertices with the insertion of one $K_0$ and $m$ $\phi$'s.
Finally $V_\phi^{(m)}$, $m=4,6,8,\dots$ denotes the number of 
 vertices with $m$ $\phi$'s and neither $J_\mu$ nor $K_0$'s.

Vertices with one $K_0$ do not contain derivatives. Vertices with
one $J_\mu$  carry one momentum while vertices with only $\phi$'s
carry two momenta.

In $D$ dimensions the superficial 
degree of divergence for the graph $G$ is 
\begin{eqnarray}
d(G) = n D - 2 I + \sum_k V_J^{(k)} + 2 \sum_j V_{\phi}^{(j)} \, .
\label{app.1}
\end{eqnarray}
Use of the Euler's relation
\begin{eqnarray}
I = n + V - 1
\label{app.2}
\end{eqnarray}
with 
\begin{eqnarray}
V =  \sum_k V_J^{(k)} + \sum_j V_{\phi}^{(j)} + \sum_l V_{K_0}^{(l)} 
\label{app.3}
\end{eqnarray}
gives
\begin{eqnarray}
d(G) = (D-2)n + 2 - \sum_k V_J^{(k)} - 2 \sum_l V_{K_0}^{(l)}
\label{app.4}
\end{eqnarray}
The above formula shows that 
at a given loop order $n$
the maximum superficial degree of divergence
in the collection of graphs with $N_J$ insertions of the composite
operator $F_\mu^a$, $N_{K_0}$ insertions of the composite operator $\phi_0$
and no $\phi$'s external legs
is obtained when the number of vertices $V_J^{(k)}$ and 
$V_{K_0}^{(l)}$ is as small as possible.

This configuration is achieved by connecting all $J_\mu$'s and all $K_0$'s
along a chain of propagators and by inserting a sufficient
number of additional propagators joining the above vertices
in such a way to generate a $n$-loop graph.
For that purpose one needs $N_J$ vertices with one $J_\mu$ and $N_{K_0}$
vertices with one $K_0$. There are $N_J + N_{K_0}$ lines in the external chain
and $n-1$ internal lines have to be added in order to get a $n$-loop graph.

The superficial degree of divergence is thus
\begin{eqnarray}
d_{\rm max}(G) & = & Dn - (2 (N_J + N_{K_0}) + 2(n-1)) + N_J \nonumber \\
          & = & (D-2)n + 2 - (N_J + 2N_{K_0}) \, .
\label{app.5}
\end{eqnarray}
$d_{\rm max}(G)<0$ if
\begin{eqnarray}
N_J + 2N_{K_0} > (D-2)n + 2 \, .
\label{app.6}
\end{eqnarray}

\end{document}